\documentclass[paper]{revtex4}

\usepackage{graphicx}
\usepackage{epsfig}
\usepackage{fancyhdr}
\usepackage{empheq}
\usepackage{mathbbol}
\usepackage{wasysym}
\usepackage{pstricks}
\usepackage{bbm}
\usepackage{bm}
\usepackage{color}
\newcommand{\be}{\begin{equation}}
\newcommand{\ee}{\end{equation}}
\newcommand{\bea}{\begin{eqnarray}}
\newcommand{\eea}{\end{eqnarray}}

\begin{document}

\title{Low Energy Neutron Production by Inverse $\beta$ decay in Metallic Hydride Surfaces}

\author{S Ciuchi$^{1}$, L Maiani$^{2}$, AD Polosa$^{2}$,  V Riquer$^{3}$, G Ruocco$^{2}$, M Vignati$^{4}$}
\affiliation{
$^{1}$Dipartimento di Scienze Fisiche e Chimiche, Universit\`a de L'Aquila, Via Vetoio, Coppito, L'Aquila,  I-67010 Italy\\
$^2$Dipartimento di Fisica, Sapienza Universit\`a di Roma, Piazzale A. Moro 5, Roma, I-00185 Italy\\
$^{3}$Fondazione TERA, Via G Puccini 11,  Novara, I-28100 Italy\\
$^{4}$INFN Sezione di Roma1, Piazzale A. Moro 5, Roma, I-00185 Italy
}

\begin{abstract}
It has been recently argued that inverse-$\beta$ nuclear transmutations might occur at an impressively high rate in a thin layer at the metallic hydride surface under specific conditions. In this note we present a calculation of the transmutation rate which shows that there is little room for such a remarkable effect. 
\end{abstract}

\maketitle
{\bf \emph{Introduction and main results}}. 
In a number of papers~\cite{widompub, widomunpub,widomext}, A. Widom and collaborators consider the intriguing possibility that low energy neutrons are produced on metallic hydride surfaces by the capture of electrons by protons and the subsequent inverse-$\beta$ reaction:
\be
\tilde{e}p\to n\nu_e
\label{invbdec}
\ee

This process cannot, of course, take place in a Hydrogen atom in vacuum, it would have a negative $Q$-value ($Q=M_p+m_e-M_n<0$). The authors advance the hypothesis that mass renormalization from the interaction of  electrons in atoms with some electromagnetic external field ({\it e.g.} due to a laser source) may increase the effective electron mass, so as to make it possible the reaction (\ref{invbdec}). The tilde sign on the electron symbol underscores that the electron bound to the proton is a `dressed' one, with an effective mass denoted by:
\be
m^*=\beta m_e
\label{meff}
\ee

A central element of the analysis of this very intriguing effect is the decay rate of the bound  $\tilde{e}p$ state which is given in\; Ref.~\cite{widomext} as\;\footnote{see Eq. (107) of Ref.~\cite{widomext}.}:
\be
\Gamma_{_\mathrm{WL}}(\tilde{e}p\to n\nu_e)=\frac{(G_F m_e^2)^2}{2\pi^2}(1+3\lambda^2) m_e(\beta-\beta_0)^2
 \label{WL}
 \ee
Natural units are used throughout, we approximate to unity the cosine of the Cabibbo angle, $\lambda=g_A/g_V\approx1.25$ is the axial to vector coupling in neutron $\beta$ decay. 

The threshold value for process (\ref{invbdec}) to occur at all is:
\be
\beta_0=\frac{M_n-M_p}{m_e}\approx 2.59
\label{betazero}
\ee

With the above numerical values and $\beta\approx 20$, the authors find: 
\be
\Gamma_{_\mathrm{WL}}(\tilde{e}p\to n\nu_e)\approx (1.86\times 10^{-3})\times (\beta-\beta_0)^2 \; \mathrm{Hz}\approx 0.56 \; \mathrm{Hz} 
\label{loro}
\ee

What is surprising in            formula (\ref{WL}) is the absence of the factor $|\psi(0)|^2$, which gives the probability for finding the electron and the proton at the same point~\footnote{A. Polosa in the seminar by Y. Srivastava in Roma, May 31, 2012.}. 
This factor is a mandatory consequence of the local Fermi lagrangian density:% ($\lambda=g_A/g_V$):

\be
{\cal L}(x)=\frac{G_F}{\sqrt{2}}\left[{\bar n}(x)\gamma_\mu (1-\lambda \gamma_5)p(x)\right] \left[{\bar \nu}(x)\gamma_\mu (1- \gamma_5)e(x)\right]
\label{fermiL}
\ee
With the electron and proton fields computed at the same point, the factor $|\psi(0)|^2$ indicates that it must be difficult for the lagrangian to induce transitions from an initial state with, say, the proton localized in our lab and the electron on Mars.
We present a calculation of the rate of (\ref{invbdec}) done in two independent ways. 

One is to consider the decay of the $\tilde{e}p$ bound states with spin $0$ and $1$, relating  the amplitude to the basic lagrangian (\ref{fermiL}) in a way analogous to the quark model calculation of pion decay made time ago by Van Royen and Weisskopf (VRW) \cite{vrw}.

The second is to use the time-honored formula:
\be
\Gamma=|\psi(0)|^2 v \; \sigma
\label{rhosigv}
\ee
where $\sigma$ is the unpolarized cross section for process (\ref {invbdec}). This is the basic formula in the theory of electron K-capture, also used \cite{landlif} to derive the decay rate of positronium from the annihilation cross section of an electron-positron pair into photons.

Either ways we find, reassuringly:

\be
\Gamma(\tilde{e}p\to n\nu_e)=|\psi(0)|^2\times  
\frac{1}{2\pi} \; (G_F\,m_e)^2 \left(1+3\lambda^2\right)\; (\beta-\beta_0)^2 
\label{resprimo}
\ee

Defining:
\be
|\psi(0)|^2 =1/R^3
\ee
the result in (\ref{WL}) is consistent with Eq. (\ref{resprimo}) for:
\be
R\approx 1/m_e \simeq 0.4 \times 10^{-2}~\mathrm{\AA}
\ee
that is the electron should be confined within its Compton radius, which is completely unrealistic.

We further set
\be
|\psi(0)|^2 = \frac{1}{\pi (a^*)^3}= \frac{\beta^3}{\pi a^3}
\ee
with $\beta$ defined in (\ref{meff}), $a$ the Bohr radius, $a=(\alpha m_e)^{-1}\simeq 0.54$~\AA~and $\alpha$ the fine structure constant, and find, finally:

\be
\Gamma(\tilde{e}p\to n\nu_e)=  
\frac{\alpha^3}{2\pi^2} \; (G_F\,m_e^2)^2 m_e \left(1+3\lambda^2\right)\; \beta^3 (\beta-\beta_0)^2 
\label{res2}
\ee

Numerically:
\be
\Gamma(\tilde{e}p\to n\nu_e @\, \beta=20)=1.8\times10^{-3}~{\rm Hz}
\ee
while, for a more moderate value, $\beta\approx2\beta_0$:
\be
\Gamma(\tilde{e}p\to n\nu_e @\, \beta=5.2)=6.9\times 10^{-7}~{\rm Hz}
\ee
The numerical value proposed in~\cite{widompub, widomunpub,widomext}, Eq.~(\ref{loro}), is obtained for $\beta = 61$ ({\it i.e.} $m^*\approx 30$ MeV). 

In the following, we give the details of the calculation of $\Gamma$. In the end, we comment on the high values of $\beta$ considered in~\cite{widompub, widomunpub,widomext}. 

{\bf \emph{Decay rates of S-wave bound states.}}
 Neutron emission can be described as arising from the decay of the two, S-wave ground states, $H_0$ and $H_1$, with total spin equal to zero and one, respectively. We may describe this decay with two phenomenological parameters, $f_{0,1}$ according to:
\be
{\cal L}_{phen}= \frac{G_F}{\sqrt{2}}\left[f_0\, {\bar n}(x)(1+\gamma_5)\nu_c\, H_0 +f_1\, {\bar n}(x)\gamma_i(1+\gamma_5)\nu_c\, H_1^i  \right]
\label{phenL}
\ee
The Fermi constant has been inserted for convenience, $H_0$ and $H_1^i$ are field operators describing the annihilation of either $H$ state and we describe the creation of the neutrino with the antineutrino field, $\nu_c$. The situation is entirely similar to the decay $\pi^- \to \mu^- \nu_c$, described by the phenomenological parameter $f_\pi$~\footnote{We understand that different flavors of neutrinos appear in $\pi$ and $H$ decays.}.

To connect to the VRW formulae, it is convenient to rewrite the Fermi lagrangian so as to have the annihilation operators for the electron and the proton in the same Dirac bilinear (similarly to the quark-antiquark fields in pion decay). This is done by first introducing the positron and antineutrino fields in (\ref{fermiL}) and then making a Fierz transformation. To wit~\footnote{when separating the term in $\gamma^0$ from the one in $\gamma^i$ we have used the relations $\gamma^0 p=p$ and $\bar n \gamma^0=\bar n$, following from the non relativistic approximation.}:
\begin{eqnarray}
&&\left[{\bar \nu}(x)\gamma_\mu (1- \gamma_5)e(x)\right]\left[{\bar n}(x)\gamma_\mu (1-\lambda \gamma_5)p(x)\right] =-\left[{\bar e_c}(x)\gamma_\mu (1+ \gamma_5)\nu_c(x)\right]\left[{\bar n}(x)\gamma_\mu (1-\lambda \gamma_5)p(x)\right] =\\
&&=\left(\frac{1+3\lambda}{2}\right)\left[{\bar e_c}(x) (1- \gamma_5)p(x)\right]\left[{\bar n}(x) (1+ \gamma_5)\nu_c(x)\right]-\left(\frac{1-\lambda}{2}\right)\left[{\bar e_c}(x)\gamma^i (1 + \gamma_5)p(x)\right]\left[{\bar n}(x) \gamma_i(1+ \gamma_5)\nu_c(x)\right]\nonumber
\label{rearr}
\end{eqnarray}

Following VRW~\cite{vrw}, we define the correctly normalized H state according to:
\bea
&& |H\rangle = (2\pi)^{3/2}\int d^3p~   f({\bm p}) \phi(r,s)(a^e_r)^\dagger({\bm p})(a^p_s)^\dagger(-{\bm p})|0\rangle
\eea
r and s are spin indices, $\phi(r,s)$ are the Clebsch-Gordan coefficients appropriate to the spin of $H$, $(a^{p, e})^\dagger$ are the proton and electron creation operators and $f({\bm p})$ the momentum space wave function, with:
\be
\int  d^3p~   |f({\bm p})|^2 =1
\ee

The parameter $f_0$ is obtained by comparing (\ref{phenL}) with (\ref{rearr}), e.g.:
\be
f_0 \langle 0|H_0|H_0\rangle =\left(\frac{1+3\lambda}{2}\right)\langle0|\left[{\bar e_c}(0) (1- \gamma_5)p(0)\right] |H_0\rangle
\ee
Expanding the fields in normal modes, spinors and gamma matrices combine with the appropriate Clebsches, and we obtain (the $\sqrt{2}$ factor comes from the Clebsch, the overall signs of $f_{0,1}$ are irrelevant):
\be
f_0=\left(\frac{1+3\lambda}{2}\right)\;\sqrt{2}\;\frac{1}{(2\pi)^{3/2}}\int  d^3p~   f({\bm p})=\left(\frac{1+3\lambda}{2}\right)\;\sqrt{2}\;\psi(0)
\ee
with the x-space wave function given by:
\be
\psi({\bm x})=\frac{1}{(2\pi)^{3/2}}\int d^3p~e^{-i{\bm p}\cdot {\bm x}}f({\bm p})
\ee
Similarly:
\be
f_1=-\left(\frac{1-\lambda}{2}\right)\;\sqrt{2}\;\psi(0)
\ee

Direct calculation of the spin-averaged rates gives:
\bea
&&\Gamma (H_0 \to n+\nu) =4\;\frac{(G_F m_e)^2}{2\pi} ~\frac{(1+3\lambda)^2}{4}|\psi(0)|^2 (\beta-\beta_0)^2 \nonumber \\
&&\Gamma (H_1 \to n+\nu) =4\;\frac{(G_F m_e)^2}{2\pi} ~\frac{(1-\lambda)^2}{4}|\psi(0)|^2 (\beta-\beta_0)^2
\label{decrates}
\eea

and the inclusive rate is obtained from:
\be
\Gamma(\tilde{e}p\to n\nu_e)=\frac{1}{4}\left[\Gamma (H_0 \to n+\nu)+3 \Gamma (H_1 \to n+\nu)\right]=|\psi(0)|^2~\frac{(G_F m_e)^2}{2\pi}(1+3\lambda^2) (\beta-\beta_0)^2
\ee
as anticipated in (\ref{resprimo}).

{\bf \emph{ Inclusive rate from the unpolarized cross section.}}
The cross section for $e^- +p\to \nu +n $ is~\cite{marshferm}:
\be
\sigma(e^- +p\to \nu +n)
=\frac{(G_F m_e)^2}{2\pi}~\left(1+3\lambda^2\right)~\frac{(\beta-\beta_0)^2}{v_e}
\ee
%where $\Delta=M_n-M_p\simeq1.29$~MeV. 
The transmutation rate thus obtained via Eq. (\ref{rhosigv}) confirms the result in (\ref{resprimo}).

{\bf \emph{The value of $\beta$.}}
Values of $\beta$ of the order or even larger than twenty are certainly unusual in condensed matter physics, expecially for bound electrons.
An estimate of $\beta$ is given in~\cite{widompub,widomunpub,widomext}
\be
\beta=\sqrt{1+A\frac{\overline{|{\bm u}|}^2}{a^2}}
\label{beta}
\ee 
where  $A$ is given in terms of the plasma frequency $\Omega_p$ of the protons:
\begin{equation}
A = \frac{M}{m^2}\; (M \Omega^2_p) a^2
\label{freq}
\end{equation}
and $\sqrt{\overline{|{\bm u}|}^2}$ is the r.m.s. displacement of the protons.

Current values of $\Omega_p$  are estimated to be of order $\Omega_p\approx 0.1$~eV~\cite{plasma}. With this value of $\Omega_p$ and $\sqrt{\overline{|{\bm u}|}^2/ a^2}\approx 4.2$~\cite{widompub,widomunpub,widomext}:
\be
\beta(\Omega_p= 0.1)=1.01
\ee
The value  $\Omega_p= 0.8$~eV  used in~\cite{widompub,widomunpub,widomext} leads to:
\be
\beta(\Omega_p= 0.8)=1.8
\ee
considerably lower than $\beta\approx 20$ and in any case below  threshold for nuclear transmutation to occur.

{\bf \emph{Conclusions.}}
A correct calculation gives a neutron production rate from (\ref{invbdec}) about $300$ times smaller than what estimated in ~\cite{widompub, widomunpub,widomext}, for the value of the mass renormalization factor $\beta\approx20$ considered there. In turn, it is questionable that values of $\beta$ can be realized, in particular for bound electrons, so large as to give rise to useful nuclear transmutation rates. A more detailed analysis of the attainable values of $\beta$ is needed to obtain more definite conclusions on this interesting phenomenon, should it exist at all.

\thispagestyle{fancy}

%{\bf \emph{Introduction}}. 
%{\bf \emph{The model}}. 
%{\bf \emph{Results and discussion}}. 
%{\bf \emph{Conclusions}}. 

%\begin{acknowledgments}

%\end{acknowledgments}

\bigskip % extra skip inserted

\end{document}